\documentclass[12pt]{article}
\usepackage{amsfonts,amssymb,epsfig,amsmath,times}
\usepackage{float}
\restylefloat{table}
\renewcommand{\text}[1]{#1}

\newcommand{\be}{\begin{equation}}
\newcommand{\ee}{\end{equation}}
\newcommand{\ben}{\begin{displaymath}}
\newcommand{\een}{\end{displaymath}}
\newcommand{\bea}{\begin{eqnarray}}
\newcommand{\eea}{\end{eqnarray}}
\newcommand{\bean}{\begin{eqnarray*}}
\newcommand{\eean}{\end{eqnarray*}}
\newcommand{\nn}{\nonumber \\}
\newcommand{\ba}{\begin{array}}
\newcommand{\ea}{\end{array}}
\newcommand{\bi}{\begin{itemize}}
\newcommand{\ei}{\end{itemize}}

\def\m{\mu}
\def\n{\nu}

\def\a{\alpha}

\def\g{\gamma}

\def\G{\Gamma}
\def\d{\delta}
\def\s{\sigma}
\def\e{\epsilon}

\def\otaula{\begin{tabular}}
\def\ctaula{\end{tabular}}

\renewcommand{\th}{\theta}
\def\bnum{\begin{enumerate}}
\def\enum{\end{enumerate}}

\def\CR{\mathcal{R}}
\def\CM{\mathcal{M}}

\def\s{\sigma}
\def\8M{$\CM_8$}
\def\k{\kappa}
\def\be{\begin{equation}}
\def\ee{\end{equation}}
\def\G{\Gamma}
\def\g{\gamma}
\def\ei{e^{\underline{i}}}

\def\e1{e^{\underline{1}}}
\def\1u{\underline{1}}
\def\2u{\underline{2}}

\def\0u{\underline{0}}
\def\e{\epsilon}
\def\target{$\CR^{1,1}\times \mathcal{M}_8$ }
\def\target2{$\CR^{1,1}\times \mathcal{M}_8$,}
\def\9G{\G_{\underline{9}}}

\def\a{\alpha}

\def\p{\partial}

\newcommand{\cale}{\mbox{${\cal E}$}}

\newcommand{\un}{\underline}

\def\w{\omega}

\def\1f{f_1^{1/2}}
\def\2f{f_2^{1/2}}
\def\4f{f_4^{1/2}}

\begin{document}
\begin{titlepage}

\vfill
\begin{flushright}
KIAS-P10001
\end{flushright}

\vfill

\begin{center}
   \baselineskip=16pt
   {\Large\bf $\mathcal{N} = 2$  SCFTs: An M5-brane perspective}
   \vskip 2cm
      Bin Chen$^1$, Eoin \'{O} Colg\'{a}in$^2$, Jun-Bao Wu$^2$  and Hossein Yavartanoo$^2$       \vskip .6cm
      \begin{small}
      $^1$\textit{ Department of Physics,\\
and State Key Laboratory of Nuclear Physics and Technology,\\
Peking University,\\
Beijing 100871, P.R. China}
        \end{small}\\*[.6cm]
      \begin{small}
      $^2$\textit{Korea Institute for Advanced Study, \\
        Seoul, Korea}
        \end{small}\\*[.6cm]
\end{center}

\vfill
\begin{center}
\textbf{Abstract}\end{center}

\begin{quote}
Inspired by the recently discovered holographic duality between $\mathcal{N}=2$ SCFTs and
half-BPS M-theory backgrounds, we study probe M5-branes. Though our main focus is supersymmetric M5-branes
whose worldvolume has an $AdS_n$ factor, we also consider some other configurations. Of special mention is the identification of $AdS_5$ and $AdS_3$ probes preserving supersymmetry, with only the latter supporting a self-dual field strength. \end{quote}

\vfill

\end{titlepage}

\section{Introduction}
Four-dimensional ${\cal N}=2$ supersymmetric theories are truely remarkable. Compared to ${\cal N}=4$ supersymmetric theories, which are
finite, they are much richer in physics, but yet still solvable. Especially, they provide a window to study the nonperturbative aspects
of quantum field theories and have been widely studied for more than fifteen years since Seiberg and Witten's monumental works \cite{SW1,
SW2}. However, the surprises they present to us have not come to the end.
Recent developments on four-dimensional ${\cal N}=2$ superconformal theories starting from
\cite{Gaiotto} have drawn lots of attention. This class of generalized quiver theories could be constructed
geometrically by wrapping M5
branes on Riemann surfaces with genus and punctures.  The electric-magnetic duality  and the Argyres-Seiberg duality \cite{AS} have since been generalized to
these theories. It turns out that the gauge couplings of the theory are encoded in the complex structure moduli
of the Riemann surface, including the position of the punctures.  More interestingly, it was conjectured in \cite{AGT} that the Nekrasov
partition function of these theories with $SU(2)$ gauge groups could be related to the conformal blocks and correlation functions
of the Liouville theory. Non-local operators in these four-dimensional theories
have been studied in \cite{DMO,Alday:2009fs, Drukker:2009id, Gaiotto2,
Wu:2009tq} (See also \cite{Zhou:2009nx}).

The holographic dual of these theories in the large N limit was neatly studied in \cite{MG}. The
theory on the gravity side of this AdS/CFT correspondence is
M-theory on backgrounds which are the products of $AdS_5$ spacetime and
six-dimensional internal manifolds with $SU(2) \times U(1)$ isometry. These gravity backgrounds
belong to the general geometries found in \cite{LLM}. Of particular interest among these backgrounds is the so-called Maldacena-N\'u\~nez(MN) geometry, which was first discovered in \cite{MN} by considering the IR limit of M5-branes
wrapped on a Riemann surface. In this case, the six-dimensional internal manifold is simply an $S^4$ fibered over
the Riemann surface. On the other side, the field theory corresponding to the MN geometry comes
from M5-branes wrapping the same Riemann surface. It is remarkable, that in this case, there are no punctures on the Riemann surface
and that the building block of the quiver gauge theory is a strongly coupled superconformal theory $T_N$ with three $SU(N)$
global symmetries, yet without a coupling constant. One motivation of this work is to
understand this intriguing $T_N$ theory from its gravity dual.

Probe branes play an important role in the gravity side of holographic correspondences. These branes are intrinsically
stringy but accessible.  Among other things, they could be
dual to local operators \cite{McGreevy00, Grisaru00, Hashimoto00},
loop operators \cite{Drukker, Yamaguchi:2006D5, Gomis06}, surface
operators \cite{Gomis:2007fi, Drukker:2008wr}, or domain walls
\cite{Witten:1998xy} in the field theory side of various AdS/CFT
correspondences. Adding suitable branes can also add flavor to the
field theory \cite{Karch:2002sh}. 

In this paper, we plan to start the search for interesting probe M5-branes
in the LLM geometries studied in \cite{MG}. We mainly focus on the
simplest MN background which we have rederived in the appendix to
include the fluxes. The form of the solution is \bea ds^{2}_{11} &=&
\tilde{\kappa}^{2/3} \biggl( \frac{1}{2} W^{1/3} ds^{2}_{AdS_5} +
\frac{W^{-2/3}}{4} \biggl[ W \frac{(dx^2 + dy^2)}{y^2} \nn &+& W d
\theta^2 + \cos^2 \theta (d \phi_1^2 + \sin^2 \phi_1 d \phi_2^2) + 2
\sin^2 \theta \left( d \chi + \frac{dx}{y}\right)^2\biggr] \biggr),
\\ H_{4} &=& \tilde{\kappa} \left( -\frac{1}{4 W^2} [ 3 + \cos^2
\theta ] \sin \theta \cos^2 \theta d \theta (d \chi + \frac{dx}{y})+
\frac{1}{4} \frac{\cos^{3} \theta}{W} \frac{dx dy}{y^2} \right) d^2
\Omega \nonumber,\eea where $\phi_1, \phi_2$ parameterize a
two-sphere, $x,y$ denote a hyperbolic Riemann surface $\Sigma_2$
of constant negative curvature, and $W$ is \be W = 1 + \cos^2 \th. \ee The
constant $\tilde{\kappa}$ denotes an additional scale factor that
will be accounted for by ensuring that the flux is correctly
quantised.\footnote{ It is related to the $\kappa$ in \cite{MG} by
$\tilde{\kappa} = 2^4 N \kappa$.}

This geometry may also be obtained
from the most general solutions of eleven dimensional supergravity
preserving $\mathcal{N}=2$ superconformal symmetry \cite{LLM} as
described in \cite{MG}. Although we focus on the above background, we believe our results can be
generalized to more general LLM geometries. In the literature, some BPS probe branes
have been studied in \cite{MG}, and half-BPS M2-brane dual to loop
operator have appeared in \cite{DMO}.

Starting with Killing spinors of MN geometry and kappa symmetry for
M5, one can search for BPS M5-branes. As a first step, one needs to determine the Killing spinors preserved by the MN solution. Luckily, this has already been done for the most general solution \cite{LLM}
and the Killing spinors corresponding to the analytically continued solution corresponding to MN have
appeared in \cite{DMO}. The latter appears without derivation, so in
the appendix we validate their claim by following a similar
decomposition to that appearing in \cite{LLM} (see also
\cite{Gaunt1})\footnote{We also cure some typos in \cite{DMO}.}. The result of that exercise is that the
eleven-dimensional Killing spinors of the MN solution can be
expressed in terms of $AdS_5$ ($\psi$) and $S^2$ ($\chi_{+}$)
Killing spinors as \bea \e &=& e^{\lambda/2} \psi \otimes  (1 + i
\sigma_3 \otimes \gamma_{(4)} ) \chi_{+}
 \otimes e^{- \frac{i}{2} \phi_0 \gamma_{10}} e^{i \chi/2} \epsilon_0, \nn \label{kills}
 \e^c &=& e^{\lambda/2} \psi^c
\otimes (1 - i \sigma_3 \otimes \gamma_{(4)} ) \chi_{+}
 \otimes e^{- \frac{i}{2} \phi_0 \gamma_{10}} e^{-i \chi/2} \gamma_7 \epsilon_0, \eea
where the superscript $c$ denotes the conjugate and \be
\sin\phi_0=\frac{\sqrt{2}\cos\theta}{\sqrt{W}},\quad
\cos\phi_0=-\frac{\sin\theta}{\sqrt{W}}, \quad e^{2 \lambda} =
\frac{\tilde{\kappa}^{2/3} W^{1/3}}{8}. \ee The constant spinor
$\e_0$ satisfies the following projection conditions \be
\label{projectors} \g_9 \e_0 = \e_0, \quad i \g_{78} \e_0 = \e_0.
\ee

With possible dual non-local objects in field theory in our mind, we
pay principal attention to M5-branes whose worldvolumes have $AdS_m
(2\le m\le 5)$ factors. The brane with worldvolume $AdS_5\times S^1$
has been studied previously in \cite{MG}. However, we find that
although turning on self-dual three-form field strength on the
worldvolume in a suitable way does not break supersymmetry, the
equations of motion of M5-branes will not be satisfied. This comes as some surprise as the Kaluza-Klein reduction of the two-form potential on the $S^1$($\chi$) gives rise to a $U(1)$ gauge field in $AdS_5$ corresponding to a global symmetry rotating the phase of a dual bifundamental field \cite{MG}. 

Moreover, we
find BPS M5-branes with an $AdS_3$ factor. This brane should be dual
to some two-dimensional object in the field theory side. However it
is not dual to the supersymmetric surface operator studied in
\cite{Alday:2009fs, Gaiotto2}, since this brane wraps the Riemann
surface in the six-dimensional internal space, instead of
intersecting with this Riemann surface at a point. In this case, we
find that we can turn on a suitable self-dual three-form field
strength on the worldvolume such that the BPS condition and the
equations of motion are both satisfied. We also find  BPS
M5-branes not embedded along the $AdS_5$ radial direction that satisfy the equations of motion hinting that there
should be non-BPS $AdS$ branes there also. In addition, inspired by
some probe M5-branes in $AdS_7\times S^4$ \cite{Lunin, Chen}, we
turn to searching for M5-branes in MN background with more complicated
worldvolumes. As a result, we find M5-branes with an $AdS_3\times
S^1$ and $AdS_2\times S^2$ factors which are completely embedded in the 
$AdS_5$ part of the background. We explicitly illustrate that generically these
branes are non-supersymmetric.

In the next section, we move to review the M5-brane equations of motion and BPS condition. With these tools at hand, we study various probe
M5-branes in section~\ref{probe}. In section~\ref{nonbps}, we
examine more exotic embeddings in $AdS_5$ before concluding. Some
technical details are located in the appendices.

\section{M5-brane review \label{M5}}

In this section, we review the M5-brane covariant equations of
motions in curved spacetime and discuss the condition for the
M5-brane probe to preserve supersymmetry. For earlier work on
various aspects of the M5-brane, see
\cite{Sezgin99,Sezgin97,Howe96,Chu97,Sundell97,Sorokin97} (For a
review of M-theory branes, see \cite{Berman}). This section echoes
the brief review of the M5-brane presented in
 \cite{Chen} and we refer the reader there for a further account of the M5-brane action.

Focusing solely on the bosonic components, we simply have two
equations of motion: a scalar and a tensor equation. The scalar
equation takes the form
 \be\label{scalareq}
 G^{mn}\nabla_m \cale_n^{\underline
 c}=\frac{Q}{\sqrt{-g}}\epsilon^{\m_1\cdots
 \m_6}\big(\frac{1}{6!}H^{\underline a}_{7~\m_1\cdots
 \m_6}+\frac{1}{(3!)^2}H^{\underline
 a}_{4~\m_1\m_2\m_3}H_{\m_4\m_5\m_6}\big)P_{\underline
 a}^{~\underline c}
 \ee
 and the tensor equation is of the form
 \be\label{tensoreq}
 G^{mn}\nabla_mH_{npq}=Q^{-1}(4Y-2(mY+Ym)+mYm)_{pq}.
 \ee
Here our notation is as follows: indices from the
beginning(middle) of the alphabet refer to frame(coordinate)
indices, and the underlined indices refer to target space ones. More details of our conventions may be found in the appendices. 

Appearing in the equations of motion, we have the following quantities which are defined in terms of the self-dual 3-form field strength $h$ on the M5-brane worldvolume
 \bea
 k_m^{~n}&=&h_{mpq}h^{npq}, \\
 Q&=&1-\frac{2}{3}\mbox{Tr} k^2, \\
 m_p^{~q}&=&\delta_p^{~q}-2k_p^{~q}, \\
 H_{mnp}&=&4Q^{-1}(1+2k)_m^{~q}h_{qnp}
 \eea
 Note that $h_{mnp}$ is self-dual with respect to worldvolume
 metric but not $H_{mnp}$. The
 induced metric is simply
 \be
 g_{mn}=\cale_m^{\underline a}\cale_n^{\underline b}\eta_{{\underline
 a}{\un b}}
 \ee
 where
 \be
 \cale_m^{\underline a}=\p_mz^{\underline m}E_{\underline m}^{\underline
 a}.
 \ee
Here $z^{\underline m}$ is a target spacetime coordinate, which becomes a function of worldvolume coordinate $\xi$ through the embedding, and
$E_{\underline m}^{\underline
 a}$ is the component of target space vielbein. From the induced
 metric, we can define another tensor
 \be
 G^{mn}=(1+\frac{2}{3}k^2)g^{mn}-4k^{mn}.
 \ee
 We also have
 \be
 P_{\underline a}^{~\underline c}=\delta^{\underline
 c}_{\underline a}-\cale_{\underline a}^m\cale_m^{~{\underline c}}.
 \ee

Note that in the scalar equation of motion, the covariant
derivative $\nabla_m\cale_n^{\underline c}$ involves not only the
Levi-Civita connection of the M5-brane worldvolume but also the
spin connection of the target spacetime geometry. More precisely,
one has
 \be
 \nabla_m\cale_n^{\underline c}=\p_m\cale_n^{\underline c}-\G^p_{mn}\cale_p^{\underline
 c}+\cale_m^{\underline a}\cale_n^{\underline b}\omega^{\underline c}_{{\underline a}{\underline
 b}}
 \ee
 where $\G^p_{mn}$ is the Christoffel symbol with respect to the induced worldvolume
 metric and $\omega^{\underline c}_{{\underline a}{\underline
 b}}$ is the spin connection of the background spacetime pulled back to the worldvolume.

 Moreover, there is a background 4-form field strength $H_{4~{\underline a}_1\cdots {\underline a}_4}$ and its Hodge dual
 7-form $H_{7~{\underline a}_1\cdots {\underline
 a}_7}$:
 \bea
 H_4&=&dC_3 \nn
 H_7&=&dC_6+\frac{1}{2}C_3\wedge H_4
 \eea
 The frame indices on $H_4$ and $H_7$ in the above equations (\ref{scalareq}) and (\ref{tensoreq}) have
 been converted to worldvolume indices with factors of
 $\cale_m^{\underline c}$.
  From the background fluxes, we can define
 \be
 Y_{mn}=[4\star {\underline H}-2(m\star {\underline H}+\star {\underline H}m)+m\star {\underline H}m]_{mn},
 \ee
where \be \star {\underline
H}^{mn}=\frac{1}{4!\sqrt{-g}}\epsilon^{mnpqrs}{\underline
H}_{pqrs} \ee

The field $H_{mnp}$ is defined by
 \be
 H=dA_2-{\un C}_3,
 \ee
 where $A_2$ is a 2-form gauge potential and ${\un C}_3$ is the pull-back
 of the bulk gauge potential. From its definition, $H$ satisfies the
Bianchi
 identity
 \be
\label{bianchi}
 dH=-{\underline H}_4
 \ee
 where ${\underline H}_4$ is the pull-back of the target space 4-form flux.

In general, the supersymmetric embeddings of a probe brane in a
background may be determined from the kappa-symmetry condition \be
\label{kappa} \G_{\kappa} \e = \pm \e. \ee Here, $\G_{\kappa}$
denotes the gamma matrix associated to the probe, $\epsilon$ denotes
the Killing spinor of the background and the sign accounts for the
choice between brane and anti-brane probes. The amount of unbroken
supersymmetry may be determined by keeping track of the additional
projection conditions that arise from the above equation.

Specializing to the MN background with M5-brane probes, the kappa
symmetry matrix $\G_{M5}$ may be written following
\cite{Howe:1997fb}
 \begin{equation}
 \Gamma_{M5}=\frac{1}{6!\sqrt{-g}}\epsilon^{j_1\cdots j_6}[\Gamma_{<j_1\cdots
 j_6>}+40\Gamma_{<j_1j_2j_3>}h_{j_4j_5j_6}].
 \end{equation}
Here $g$ is the determinant of the induced worldvolume metric
component, $h_{j_4j_5j_6}$ is the self-dual 3-form on the
M5-brane and $\Gamma_{<j_1\cdots j_n>}$ is defined as
 \begin{equation}
 \Gamma_{<j_1\cdots j_n>}={\cal E}^{\underline a_1}_{j_1}\cdots{\cal E}^{\underline
 a_n}_{j_n}\Gamma_{{\underline a_1}\cdots {\underline a_n}},
 \end{equation}
 where $\Gamma_{{\underline a_1}\cdots {\underline a_n}}$ is the product of the Gamma matrices
 in orthonormal frame.

We pause here to make a brief comment. Denoting the worldvolume of
the M5 by $\xi^{a}, a = 0,\cdots 5$, in the case of a simple probe
configuration, we may rewrite the above projector (\ref{kappa}) as
\be \left[ \alpha \G_{012345} + \beta (\G_{012} - \G_{345}) \right]
\epsilon = \pm \epsilon, \ee where $\alpha, \beta$ denote arbitrary
factors. Demanding it to be a projector, it is essential the left
hand side squares to unity. In that event, $\beta$ drops out
completely and $\alpha^2 = 1$, meaning that $\alpha = \pm 1$. The
implication of this observation, at least for the simple
probes considered in this paper, is that if the M5-probe is not
supersymmetric, then supersymmetry cannot be restored by introducing
$h$. So the task in the rest of the paper is pretty straightforward:
identify supersymmetric probes and then turn on $h$ to see if it
preserves supersymmetry. At each stage, it is also imperitive to
ensure that the equations of motion are satisfied.

\section{Supersymmetric probes\label{probe}}
In this section, we focus on the kappa-symmetry condition
(\ref{kappa}) and isolate probes that will preserve some
supersymmetry. We descend in dimension of the part in $AdS_5$ from
d=5 to d=2 and in each case, we enumerate the possibilities.
Throughout we differentiate between probes that are $AdS$ i.e. those
incorporating the radial direction $r$ of $AdS_5$ and those located
at a fixed $r$ . We begin by examining the $AdS$ probes.

\subsection{Supersymmetric $AdS$ probes}
In this subsection, we descend from $AdS_5$ to $AdS_2$ and identify the supersymmetric probes (if any), before examining the additional constraints coming from the equations of motion. As a warm-up, we begin
with the $AdS_5$ M5-brane probe which received some attention in \cite{MG}.\\

\noindent
$\mathbf{AdS_5}$ \textbf{probes} \\
In general, one can consider studying the probe brane with
worldvolume $AdS_5 \times \mathcal{C}$ in the MN background, where
$\mathcal{C}$ denotes a curve in the six-dimensional space transverse to $AdS_5$. We consider the $\mathcal{C}$ to be parameterised by $\sigma$ i.e $z^{\un m}(\s)$. Using $AdS_5$ Poincar\'{e} coordinates, a natural choice for the M5 embedding is
\be \xi_{0} = x_0, \quad \xi_{i} = x^{i}, \quad \xi_{4} = r, \quad \xi_{5} = \s, \ee where $i=1,2,3$.

Adopting the gauge choice $\sigma = \chi$, while permitting embeddings of the form $x \equiv x(\chi), y \equiv y(\chi)$, the kappa symmetry matrix
$\G_{M5}$ simplifies to
\be \G_{M5} = \frac{1}{\sqrt{g_{\chi \chi}}} \G_{01234} \G_{<\chi>}
,\ee
 where the induced metric component is \be g_{\chi \chi} =
\left( \frac{\tilde{\kappa}}{W} \right)^{2/3}\left( \frac{W}{4} \left(
\frac{(\p_{\chi}x)^2+(\p_{\chi}y)^2}{y^2} \right) + \frac{\sin^{2} \theta}{2}  \left( 1 +
\frac{(\p_{\chi} x)}{y} \right)^2\right), \ee and the induced gamma
matrix is \bea \G_{<\chi>} &=& 1_4 \otimes 1_2 \otimes \left( \frac{\tilde{\kappa}}{W} \right)^{1/3} \left[ \frac{\sin \th}{\sqrt{2}}
\left( 1 + \frac{(\p_{\chi} x)}{y}\right) \g_{{9}} + \frac{W^{1/2}}{2} \left(
\frac{(\p_{\chi} x)}{y} \g_{{7}} + \frac{(\p_{\chi}
y)}{y} \g_{{8}} \right) \right]. \nn \eea

Utilising $\rho_{01234} = i$, the requirement for supersymmetry
$\G_{M5} \e = \pm \e$ then reduces to \be \frac{\tilde{\kappa}^{1/3}
W^{-1/3}}{\sqrt{g_{\chi \chi}}}\left[ \frac{\sin \th}{\sqrt{2}}
\left( 1 + \frac{(\p_{\chi} x)}{y}\right) \g_{{9}} +
\frac{W^{1/2}}{2} \frac{(\p_{\chi} x)}{y} \g_{{7}} +
\frac{W^{1/2}}{2} \frac{(\p_{\chi} y)}{y} \g_{{8}}\right] \e_0 = \pm
\e_0,  \ee provided $\phi_0 = \pi$. This means that $\th =
\tfrac{\pi}{2}$ and $W=1$. In addition, we rquire the probe to be
located at a point on the Riemann surface: \be \p_{\chi} x =
\p_{\chi} y = 0, \ee so that the terms proportional to $\g_7$ and
$\g_8$ disappear. The projection condition $\g_9 \e_0 = \e_0$ also
singles out the positive sign above indicating that the probe is an
M5-brane as opposed to an anti-M5-brane.

Therefore, the curve $\mathcal{C}$ is exclusively along the
$\chi$-direction. As noted in \cite{MG}, the $\theta =
\tfrac{\pi}{2}$ condition corresponds to the $S^2$ shrinking, so the
superconformal symmetry $SU(2) \times U(1)$ symmetry of the
background is preserved by this probe. Also, no supersymmetry is
broken by this probe.

Now that we have a supersymmetric probe, we may inquire whether it
is possible to turn on self-dual $h$. As explained earlier, this
problem reduces to ensuring \be a (\G_{012} - \G_{349}) \epsilon =
0, \ee where we have defined \be h = \frac{a}{2}(\cale^{012} +
\cale^{349}).\ee Again using the decomposition (\ref{gamma_dec}) and the
relationship $\rho_{01234} = i \Rightarrow \rho_{34} = i
\rho_{012}$, then it is possible to show that this condition is
satisfied.

Having verified the kappa-symmetry condition is satisfied, it remains to show that the equations of motion are satisfied.
The induced metric may be written \bea ds^{2}_{ind} &=&
\frac{\tilde{\kappa}^{2/3}}{2} \left[ \frac{dx_{\m} dx^{\m} + dr^2}{r^2} + d \chi^2 \right],\eea where we have used Poincar\'{e} coordinates.

The RHS of the  tensor equation (\ref{tensoreq}) is zero as the
background 4-form flux does not pull back to the M5 worldvolume. The
tensor equation is then simply \be  \quad G^{mn}\nabla_mH_{npq} = 0. \ee
As for scalar equation, the RHS vanishes trivially when $c\ne 10$.
For the case with $c=10$, the RHS is non-vanishing for general
$\theta$ due to the $vol_{AdS_5}\wedge d\theta \wedge d\chi$ part of
$H_7$. However, the coefficient is proportional to $\cos\th$, so it
vanishes when it gets pulled back to the worldvolume at $\th=\tfrac{\pi}{2}$. So, neglecting this case, the scalar equation is simply
\be G^{mn}\nabla_m \cale_n^{\underline
 c}= 0.
\ee
This equation is quickly confirmed to be satisfied as it only
has one non-trivial component: \bea \nabla_{r}
\mathcal{E}^{\underline{4}}_{r} &=& \p_r
\frac{\tilde{\kappa}^{1/3}}{\sqrt{2} r} +
\frac{\tilde{\kappa}^{1/3}}{\sqrt{2} r^2 }
 = 0.  \eea

The ansatz we consider for $h$ is \bea h &=& \frac{a}{2} \left(
\cale^{012} + \cale^{349} \right) \nn &=& \frac{a \tilde{\kappa}}{4
\sqrt{2}} \left( \frac{1}{r^3} dt dx_1 dx_2 + \frac{1}{r^2} dx^3 dr
d \chi \right),\eea where $a$ is a function of r. Following the
treatment in \cite{Chen}, $H$ may be expressed as \be H =
\frac{a \tilde{\kappa}}{(1+a^2)\sqrt{2}} \frac{1}{r^3} dt dx_1 dx_2
+ \frac{a \tilde{\kappa}}{(1-a^2)\sqrt{2}}\frac{1}{r^2} dx^3 dr d
\chi. \ee As the background 4-form flux doesn't pull back, the
Bianchi identity (\ref{bianchi}) is simply $d H =0$. This means
that \be \frac{a}{(1+a^2)r^3} = \mbox{constant}.\ee Switching the
location of $dr$ in the flux ansatz above would make this \be
\frac{a}{(1-a^2)r^2} = \mbox{constant}. \ee Once the Bianchi is
satisfied, one may return to the tensor equation. Here $G^{mn}$ is
diagonal and the only term of interest is $\nabla_r H_{rx_3\chi}$
which is not zero unless $a$ is a constant. So, we reach a contradiction and the conclusion is that there is no
supersymmetric $AdS_5$ M5-brane probe with self-dual 3-form $h$.\\

\noindent
$\mathbf{AdS_4}$ \textbf{probes} \\
For $AdS_4$ probes, a quick look at appendix B reveals that we must mix the spinor $\psi$ with its conjugate $\psi^{c}$. This is because $\rho_{0124} \eta_{+} $ and $\eta_{-}$ both have the same eigenvalue under $\rho^{4}$. Thus, we consider
\be
\label{ads4_proj}
\rho_{0124} \psi = c \psi^{c},
\ee
where $c$ is a constant.
The overall effect of this mixing is that the MN Killing spinor gets related to its conjugate through the kappa-symmetry condition.

Adopting the M5 embedding \be \xi_0 = x_0, \quad \xi_i = x_i
(i=1,2),  \quad \xi_3 = r,  \quad \{ \xi_4, \xi_5 \} \subset M_6,
\ee where $M_6$ denotes the space transverse to $AdS_5$, the kappa-symmetry condition may be re-written as \bea c \g_7
\frac{\G_{(2)}}{\sqrt{g_2}} (1+ i \s_3 \otimes \g_{(4)}) \chi_{+}
e^{- \frac{i}{2} \phi_0 \gamma_{10}} e^{i \chi} \epsilon_0 = (1+ i \s_3 \otimes
\g_{(4)}) \chi_{+} e^{+ \frac{i}{2} \phi_0 \gamma_{10}} \epsilon_0,
\eea where we have multiplied across by $e^{i \chi/2}  \g_7 $ and used $\G_{(2)}
\equiv \G_{<\xi_4\xi_5>}$. One may quickly recognize that a
necessary conditions for supersymmetry are \bea \label{cond1}
\left[\g_7 {\G_{(2)}}, \s_3 \otimes \g_{(4)} \right] &=& 0, \\
\label{cond2} \left\{ \g_7 {\G_{(2)}}, \g_{10} \right\} &=& 0. \eea
The latter condition may be ignored if $\phi_0 = \pi$. The
directions transverse to the $AdS_5$ space are the product of a
two-sphere with a four-dimensional space $M_6 = S^2 \times M_4$.
Thus, in general, $\G_{(2)}$ can be a linear combination of two
anti-symmeterised gamma matrices, either along $S^2$, along $S^1
\subset S^2$ with a direction in $M_4$, or along $M_4$. The three
possibilities for $\g_7 \G_{(2)}$ are, respectively \be i \s_3
\otimes \g_7, \quad \s_i \otimes \g_7 \g_{(4)} \g_{\mu}, \quad 1_2
\otimes \g_7 \g_{\m \n}, \ee where $i=1,2$ and $\m,\n=7,8,9,10$. All
three choices fail to satisfy (\ref{cond1}), so there is no
supersymmetric probe with this embedding.\\

\noindent
$\mathbf{AdS_3}$ \textbf{probes} \\
For $AdS_3$ probes there is no mixing required between the conjugate MN Killing spinors, so for simplicity, we simply use $\e = \psi \otimes \xi$ and ignore the conjugate. Referring to (\ref{poincare}) and (\ref{sconf}), this allows us to identify $\eta_{+}$ as Poincar\'{e} and $\eta_{-}$ as superconformal Killing spinors, respectively.

After some trial and error, one identifies only one promising candidate embedding
\be
\xi_0 = x_0, \quad \xi_1 = x_1, \quad \xi_2 = r, \quad \{\xi_3, \xi_4 \} \subset \Sigma_2, \quad \xi_5 = \chi.
\ee

Using the AdS projection\footnote{The sign choice here identifies the probe as an M5. An anti-M5 maybe considered by changing sign. }\be \label{ads3_proj} \rho_{014} \psi = -\psi,  \ee the kappa-symmetry condition becomes
\be
\frac{\G_{<xy\chi>}}{\sqrt{g_3}} \e_0 = \g_{789} \e_0 = -i\e_0,
\ee
where the probe is required to be at $\th = \tfrac{\pi}{2}$ to preserve supersymmetry.
The AdS projector $\rho_{014} \psi = - \psi$ means that $\rho_{01} \eta_{+} = -\eta_{+}$ and
 $\rho_{01} \eta_{-} =  \eta_{-}$ provided the probe is located at $x_2=x_3 = 0$ in $AdS_5$. So we can preserve 4 Poincar\'{e} and 4 superconformal supersymmetries. As in the case of $AdS_5$, the $SU(2) \times U(1)$ superconformal symmetry is preserved by this probe.

For this supersymmetric probe, the equations of motion can be shown to be satisfied. Here, the induced metric is now
\be
 ds^{2}_{ind} =
\frac{\tilde{\kappa}^{2/3}}{2} \left[ \frac{-dx_0^2 + dx_1^2 + dr^2}{r^2} + \frac{1}{2}\frac{(dx^2 + dy^2)}{y^2} +  \left(d \chi + \frac{dx}{y} \right)^2 \right].
\ee
The scalar equation (\ref{scalareq}) reduces to two non-trivial components
\be
G^{rr}\nabla_{r} \cale_{r}^{\un 4} = G^{yy}\nabla_{y} \cale_{y}^{\un 8} = 0,
\ee
which may be easily verified to hold. The ansatz for $h$
\be
h = \frac{a \tilde{\kappa}}{4 \sqrt{2}} \left( \frac{dx_0 \wedge dx_1 \wedge dr}{r^3} + \frac{dx \wedge dy \wedge d \chi}{2y^2} \right),
\ee
ensures that both the tensor equation (\ref{tensoreq}) and  Bianchi (\ref{bianchi}) are satisfied when $a$ is constant. \\

\noindent
$\mathbf{AdS_2}$ \textbf{probes} \\
As in the case of $AdS_4$ treated earlier, here we also need to mix MN Killing spinor conjugates. Again, we choose
\be
\rho_{04} \psi = c \psi^{c},
\ee
with $c$ constant. We also adopt the M5 embedding
\be
\xi_0 = x_0, \quad \xi_1 =r, \quad \{ \xi_2,...,\xi_5 \} \subset M_6.
\ee

In similar fashion to steps taken before, the kappa-symmetry condition now reads
\be c \g_7 \frac{\G_{(4)}}{\sqrt{g_4}} (1+ i \s_3 \otimes \g_{(4)})
\chi_{+} e^{- \frac{i}{2} \phi_0 \gamma_{10}} e^{i \chi} \epsilon_0 = (1+ i
\s_3 \otimes \g_{(4)}) \chi_{+} e^{+ \frac{i}{2} \phi_0 \gamma_{10}}
\epsilon_0, \ee where now $\g_7 \G_{(4)} \equiv \g_7
\G_{<\xi_2...\xi_5>}$ is a linear combination of the building
blocks \be i \s_3 \otimes \g_7 \g_{\m \nu}, \quad \s_i \otimes  \g_7
\g_{(4)} \g_{\mu \n \rho}, \quad  1_2 \otimes \g_7 \g_{(4)}. \ee
These correspond to the probe wrapping $S^2$, wrapping $S^1 \subset
S^2$, and the probe not wrapping $S^2$, respectively. Before we can
even consider talking about projection conditions, a necessary
condition for supersymmetry is that $\g_7 \G_{(4)}$ commutes with
$\s_3 \otimes \g_{(4)}$. In much the same way as for $AdS_2$, this
condition is not satisfied, thus ruling out the possibility of a simple
supersymmetric $AdS_2$.

\subsection{Other supersymmetric probes}
In this section, we repeat the steps of the last section for probes
at a fixed value of $r$. We catalogue the possibilities below. In
general, one may consider change the embedding along a spatial
direction of $AdS_5$ into the radial direction $r$, getting new
M5-brane configurations. We anticipate that those non-supersymmetric
configurations are also solutions to the equations of motion. For
$M_4$, these branes should correspond to domain walls on the field
theory side. It would be interesting to eventually study how the
gauge theory changes when
one crosses the domain wall.\\

\noindent
$\mathbf{M_4}$ \textbf{probes} \\
Here again we have no mixing between the MN Killing spinor and its conjugate, so we opt to just work with $\e = \psi \otimes \xi$.
Introducing the projector
\be \label{m13_proj}
\rho_{0123} \psi = \pm i \psi, \ee we adopt the following embedding for the M5-brane
\be
\xi_0 = x_0, \quad \xi_i = x_i~(i=1,2,3), \quad \{ \xi_4,\xi_5 \} \subset M_6.
\ee
After some preliminary trial and error using the background
projectors (\ref{kills}), one finds that there are two promising
candidates for embeddings: \be \{ \xi_4,\xi_5 \} \subset \Sigma_2 \mbox{ and } \{ \xi_4,\xi_5 \} \subset S^2. \ee

For the first embedding, we take $\chi$ to be a function
of $x,y$ i.e. $\chi \equiv \chi(x,y)$. We also neglect $\theta$ as
there is no $\g_{10}$ projector acting on the MN Killing spinors.
Making use of $\rho_{0123} \psi = i \psi$, the kappa symmetry
condition becomes \be \frac{i \G_{<xy>} }{\sqrt{g_2}} \e_0 = \e_0, \ee
where $g_2$ denotes the induced metric. The induced gamma matrices
are \bea \label{ads4_ind} \G_{<x>} &=& \tilde{\k}^{1/3} \left[ \frac{ W^{1/6}}{2y} \g_7
+ \frac{\sin \th W^{-1/3}}{\sqrt{2} y} \g_9 + \p_{x} \chi \frac{\sin
\th W^{-1/3}}{\sqrt{2}} \g_9\right], \nn \G_{<y>} &=& \tilde{\k}^{1/3}
\left[ \frac{ W^{1/6}}{2y} \g_8 + \p_{x} \chi \frac{\sin \th
W^{-1/3}}{\sqrt{2}} \g_9 \right]. \eea Setting $\th = 0$, we find
that this configuration is supersymmetric with 4 Poincar\'{e}
supersymmetries preserved. One may also switch on $h$-field of the form \be h
= \frac{a}{2} (\cale^{012} + \cale^{378}). \ee

In the second case, the kappa symmetry condition becomes \be
\sigma_3 \chi_{+} =  \pm \chi_{+}. \ee On top of the projector
(\ref{m13_proj}), this breaks supersymmetry further. So we are left
with 2 Poincar\'{e} supersymmetries. There is no condition on
$\theta$ and provided we avoid $\th = \tfrac{\pi}{2}$ where the
$S^2$ shrinks. It is easy to show that the above projector also
supports a self-dual $h$-field like
 \be h = \frac{a}{2}
(\cale^{012} + \cale^{378}).\ee The equations of motion give more
constraints. One of the scalar equations require $\theta$ should be
zero, for both vanishing $h$ and the non-zero $h$-field given 
above.

\vspace{3mm}
\noindent
$\mathbf{M_3}$ \textbf{probes} \\
We find there is no simple supersymmetric probe obtained by mixing $\psi$ with its conjugate $\psi^c$. The difficulties presented in treating the relevant factor $e^{i \chi}$ will also be found in an $M_1$ probe, which we choose not to pursue.

%In contrast to the previous section where it was not necessary to mix $\psi$ with its conjugate $\psi^c$, here we have to choose the projector
%\be
%\rho_{012} \psi = c \psi^c.
%\ee
%After a little work, one reaches the conclusion that the only candidate embedding is
%\be
% \xi_{0}= x_0, \quad\xi_{i} = x_{i} (i=1,2), \quad \xi_3 = S^1 \subset S^2, \quad \{\xi_4, \xi_5\} \subset \{\Sigma_2, \chi\}.
%\ee
%There are now two choices. Adopting the embedding $\xi_4 = x, \xi_5 = \chi$ and $y$ constant, the kappa-symmetry condition for an M5-probe reduces to the two projectors
%\be
%i c \s_2 \chi_+ = \pm \chi_+, \quad
%\g_9 \e_0 =  \e_0,
%\ee
%while the second choice $\xi_4 = y, \xi_5 = \chi$ with $x$ constant leads to the following projectors
%\be
%c \s_2 \chi_+ = \pm \chi_+, \quad, i \g_{78} \e_0 = \e_0.
%%\ee
%In each case there is no constraint on $theta$ provided none of the cycles shrink and $c$ must be chosen appropriately. The presence of the two additional projectors means that the supersymmetry of the background is broken to a quarter.
\vspace{3mm}
\noindent
$\mathbf{M_2}$ \textbf{probes} \\
As in the case of $M_4$, we confine our attention to $\e = \psi \otimes \xi$. Introducing the projection condition
\be
\rho_{01} \psi = \pm \psi,
\ee
we consider an embedding of the form
\be
 \xi_{0}= x_0, \quad\xi_{1} = x_{1} \quad \{ \xi_2, \xi_3 \} \subset S^2, \quad \{\xi_4, \xi_5\} \subset \Sigma_2.
\ee This embedding preserves supersymmetry provided we allow for the
additional projector \be \s_3 \chi_+ = \pm \chi_+. \ee As $i
\g_{78}$ commutes with $\g_{10}$, supersymmetry does not pick out a
specific angle for $\theta$ provided we avoid the $S^2$ shrinking.
In total 2 Poincar\'{e} supersymmetries are preserved. This ansatz
satisfies the M5 equations of motion when $\th=0$. We can get
non-BPS brane with an $AdS_2$ factor from the brane with an ${\cal
M}_2$ factor. This non-BPS brane should be dual to some
one-dimensional object in the field theory side.

\section{More non-supersymmetric probes\label{nonbps}}
Inspired by some M5-branes in $AdS_7\times S^4$ \cite{Lunin,
Chen}, in this section we consider fibrations of $AdS_5$ that may
give rise to loop operators ($AdS_2 \times S^1 \subset AdS_5, ~AdS_2
\times S^2 \subset AdS_5$) and surface operators ($AdS_3 \times S^1
\subset AdS_5$). We begin by analysing the equations of motion for
the $AdS_3$ M5-brane probes which turn out to be instructive in
making comments about the $AdS_2$ case.

We begin with an M5 whose worldvolume is $AdS_3\times S^1\times
\Sigma_2$ with $AdS_3\times S^1$ in $AdS_5$.
 We write
$ds^2_{AdS_5}$ as:
\be ds^2_{AdS_5}=\cosh^2\rho(-\cosh^2\zeta
d\tau^2+d\zeta^2+\sinh^2\zeta d\varphi^2)+d\rho^2+\sinh^2\rho
d\alpha^2.
\ee
The veilbein of the background geometry is:
\bea
{E}^{\underline{0}}&=&\frac{\tilde\kappa^{1/3}W^{1/6}}{\sqrt{2}}\cosh\rho\cosh\zeta
d\tau, \hspace{10mm}
{E}^{\underline{1}}=\frac{\tilde\kappa^{1/3}W^{1/6}}{\sqrt{2}}\cosh\rho
d\zeta, \\
{E}^{\underline{2}}&=&\frac{\tilde\kappa^{1/3}W^{1/6}}{\sqrt{2}}\cosh\rho\sinh\zeta
d\varphi,\hspace{10mm}
{E}^{\underline{3}}=\frac{\tilde\kappa^{1/3}W^{1/6}}{\sqrt{2}}d\rho,\\
{E}^{\underline{4}}&=&\frac{\tilde\kappa^{1/3}W^{1/6}}{\sqrt{2}}\sinh\rho
d\alpha, \hspace{21mm}
{E}^{\underline{5}}=\frac{\tilde\kappa^{1/3}W^{-1/3}\cos\th}2
d\phi_1, \\
{E}^{\underline{6}}&=&\frac{\tilde\kappa^{1/3}W^{-1/3}\cos\th\sin\phi_1}2d\phi_2, \hspace{9mm}
{E}^{\underline{7}}=\frac{\tilde\kappa^{1/3}W^{1/6}}2\frac{dx}y,\\
{E}^{\underline{8}}&=&\frac{\tilde\kappa^{1/3}W^{1/6}}2\frac{dy}y,\, \hspace{31mm}
{E}^{\underline{9}}=\frac{\tilde\kappa^{1/3}W^{-1/3}\sin\th}{\sqrt{2}}(d\chi+\frac{dx}y),\\
{E}^{\underline{10}}&=&\frac{\tilde\kappa^{1/3}W^{1/6}}2d\th.
\eea

The ansatz of the M5-brane is: \be \xi_0=\tau, \xi_1=\zeta,
\xi_2=\varphi, \ee \be \xi_3=\alpha, \xi_4=x, \xi_5=y,\ee
\be\rho=\rho_0, \th=\th_0.\ee
 The induced metric is: \bea d\tilde
s^2&=&\tilde\kappa^{2/3}\frac12W_0^{1/3}(-\cosh^2\rho_0\cosh^2\zeta
d\tau^2+\cosh^2\rho_0 d\zeta^2+\cosh^2\rho_0\sinh^2\zeta
d\varphi^2)\nonumber\\
&+&\tilde\kappa^{2/3}(\frac12W_0^{1/3}\sinh^2\rho_0d\alpha^2+\frac{W_0^{1/3}(dx^2+dy^2)}{4y^2}
+\frac1{2W_0^{2/3}}\sin^2\th_0 \frac{dx^2}{y^2}),\label{induced}
\eea where \be W_0=1+\cos^2\th_0.\ee

The ansatz for $h$ is: \bea
h&=&\frac{a}2\kappa(\frac{W_0^{1/2}}{2\sqrt{2}}\cosh^3\rho_0\sinh\zeta
\cosh\zeta\d\tau d\zeta d\varphi
+\frac{W_0^{1/2}}{4\sqrt{2}}\frac{\sinh\rho_0}{y^2}d\alpha dxdy).
\eea

From this we can get: \be k_m^{\,\,n}=\left(\begin{array}{cc}
-\frac{a^2}2I_3&0\\
0&\frac{a^2}2I_3
\end{array}\right), \ee
\be Trk^2=\frac32a^4, Q=1-a^4,\ee \bea
H&=&2a\kappa^{-1}(\frac{W_0^{1/2}}{2\sqrt{2}(1+a^2)}\cosh^3\rho_0\sinh\zeta\cosh\zeta
d\tau\w d\zeta \w
d\varphi\nn\\
&+&\frac{W_0^{1/2}}{4\sqrt{2}(1-a^2)}\frac{\sinh\rho_0}{y^2}d\alpha
dxdy) \eea

From $\underline{H}_4=0$, we get that $dH=\underline{H}_4$ is
satisfied if $a$ is constant.

The results for $G^{mn}$ is: \be G^{\tau\tau}=(1+a^2)^2g^{\tau\tau},
G^{\zeta\zeta}=(1+a^2)^2g^{\zeta\zeta},
G^{\varphi\varphi}=(1+a^2)^2g^{\varphi\varphi}\ee \be
G^{\a\a}=(1-a^2)^2g^{\a\a}, G^{xx}=(1-a^2)^2g^{xx},
G^{yy}=(1-a^2)^2g^{yy}.\ee The tensor equations now become: \be
G^{mn}\nabla_mH_{npq}=0.\ee The above ansatz satisfy these equations
when $\th_0=0$.

The scalar equations become: \be G^{mn}\nabla_m{\cal
E}^{\underline{c}}_n=0, \ee for $c\ne 3$. When $\theta_0=0$, The
equations for $c\ne 3$ are satisfied automatically. For the equation
with $c=3$, the RHS is non zero due to $H_7$. Among three terms in
$H_4$, only the following contributes: \be
-\frac{2\sqrt{2}(3+\cos^2\th)}{\tilde\kappa^{1/3}W^{7/6}}
e^{\underline{789(10)}}, \ee whose Hodge dual is: \be
-\frac{2\sqrt{2}(3+\cos^2\th)}{\tilde\kappa^{1/3}W^{7/6}}e^{\underline{0123456}},\ee
The  scalar equation with $c=3$ is: \be
\frac{\sqrt{2}}{\tilde\kappa^{1/3}W_0^{1/6}}(3(1+a^2)^2\frac{\sinh\rho}{\cosh\rho}+
(1-a^2)^2\frac{\cosh\rho}{\sinh\rho})=
\frac{2\sqrt{2}(3+\cos^2\th_0)(1-a^4)}{\tilde\kappa^{1/3}W_0^{7/6}}\ee
By using $\theta_0=0$, we get \be
3(1+a^2)^2\frac{\sinh\rho}{\cosh\rho}+
(1-a^2)^2\frac{\cosh\rho}{\sinh\rho}=4(1-a^4). \ee By the mean value
inequality, the absolute value of the LHS is not less than \be
2\sqrt{3}|1-a^4|,\ee so this equation has real root when $|a|\le 1$.

If we consider M5 with $AdS_2\times S^2\times\Sigma_2$ with
$AdS_2\times S^2$ inside $AdS_5$, it seems that the only change is
the scalar equation with $c=3$, we get
 \be
2(1+a^2)^2\frac{\sinh\rho}{\cosh\rho}+
2(1-a^2)^2\frac{\cosh\rho}{\sinh\rho}=4(1-a^4). \ee We have similar
results as above. We have also verified that there are no M5-brane
solutions with factor $AdS_2 \times S^1 \subset AdS_5$. 

\subsection{Supersymmetry} In this subsection we identify the conditions
for supersymmetry to be preserved by the above probes. We begin with $AdS_3$ by solving the Killing spinor equation for an $AdS_3 \times S^1$ fibration of $AdS_5$. The result of the analysis stipulates that the probe has to be located at $\rho = \infty$ for supersymmetry to be preserved. Therefore, we validate our claim that the above probe located at finite $\rho$ is non-supersymmetric. We also sketch the calculation for $AdS_2 \times S^2$.

In each case we employ the same fibration with vielbein
\be E^{a} = \cosh \rho \; \bar{E}^a_{AdS_n}, \quad
e^{n} = d \rho, \quad e^{\alpha} = \sinh \rho \bar{E}^{\a}_{S^{4-n}}, \ee where
$a=0,..,n-1$ and $\a = n+1,..,4$. .

For $AdS_3 \times S^1$, we introduce the following decomposition for the $AdS_5$
gamma matrices \be \rho_{a} = \tau_a \otimes \s_3, \quad  \rho_3 = 1
\otimes \s_1, \quad \rho_4 = 1 \otimes \s_2, \ee where \be
\{\tau_a,\tau_b\}= 2\eta^{ab}, \ee and $\s_i$ denote the Pauli
matrices. By further writing the $AdS_5$ spinor, $ \psi$ as the
product $\psi \equiv \chi \otimes \xi$, the Killing spinor equation
on $AdS_5$ become \bea
\label{eq1} \frac{1}{\cosh \rho} \xi + \frac{\sinh \rho}{\cosh \rho} i \s_2 \xi &=& \s_3 \xi, \\
\label{eq2} \partial_{\rho} \xi &=& \frac{\s_1}{2} \xi, \\
\label{eq3} \partial_{\alpha} \xi + \frac{i \s_3}{2} \cosh \rho \xi
&=& \frac{\s_2}{2} \sinh \rho \xi. \eea Here, we have used the
Killing spinor equation on $AdS_3$: $\nabla_{a} \chi =
\tfrac{\tau_a}{2} \chi$. Writing the two-dimensional complex spinor
as \be \xi = \left( \begin{array}{c} \xi_1 \\ \xi_2 \end{array}
\right), \ee (\ref{eq1}) tells us that the two components are not
independent: \be \sinh \left( \frac{\rho}{2}\right) \xi_1 = \cosh
\left( \frac{\rho}{2} \right) \xi_2. \ee Then solving (\ref{eq2})
and (\ref{eq3}), we find that the solution is of the final form for
the $AdS_5$ Killing spinor is \be \psi \equiv \chi \otimes e^{-i
\alpha/2} \left( \begin{array}{c} \cosh \left( \frac{\rho}{2}
\right) \\ \sinh \left( \frac{\rho}{2}\right)  \end{array}\right), 
\ee where $\alpha$ denotes one of the angles of the $S^2$.  

We can now determine the supersymmetry condition on a flat probe
in the $AdS_3 \times S^1$ directions. The projection condition
$\rho^{0124} \psi =
 i \psi$ implies \be \sinh \rho =  \cosh
\rho. \ee So, supersymmetry can be preserved at $\rho =  \infty$.

For $AdS_2 \times S^2$ the calculation runs as follows. We introduce the gamma matrix decomposition
\be \rho_{a} = \tau_a \otimes 1, \quad \rho_{2} = \tau_3 \otimes \s_3, \quad \rho_{\a} = \tau_3 \otimes \sigma_{\alpha}, \ee
and the following decomposition for the spinor
\be
\psi = f^{AB} \chi_{A} \otimes \xi_{B},
\ee
where the $AdS_2$ spinor $\chi_{A}$ and the $S^2$ spinor $\xi_{B}$ satisfy the Killing spinor equations
\bea
\nabla_{a} \chi_{\pm} &=& \pm \tfrac{1}{2} \tau_3 \tau_{a} \chi_{\pm} , \nn
\nabla_{\alpha} \xi_{\pm} &=& \pm \tfrac{1}{2} \s_a \xi_{\pm}.
\eea
The components of the respective spinors are also related via
\be
\tau_{3} \chi_{+} = \chi_{-}, \quad \s_{3} \xi_{+} = \xi_{-}.
\ee
With this set-up, placing $\psi$ directly into the $AdS_5$ Killing spinor equation, one arrives at the following form of the Killing spinor
\be
\psi = \cosh (\tfrac{\rho}{2}) \chi_{+} \otimes \xi_{+} + \sinh (\tfrac{\rho}{2}) \chi_{-} \otimes \xi_{-},
\ee
with $f^{+-} = f^{-+} = 0$. One can then readily verify that the probe projector $\rho_{0134} \psi = i \psi$ can only be solved when $\rho = \infty$.

\section{Conclusion}

In this paper, motivated by the recent advances in our understanding of $\mathcal{N} = 2$ SCFTs,  we have attempted to identify simple probe M5-branes in the MN
background preserving supersymmetry. In addition to the known $AdS_5 \times S^1$ probe, our analysis identified an $AdS_3 \times \Sigma_2 \times S^1$ counterpart embedding that breaks supersymmetry further. As the M5 also wraps the Riemann surface in this case, its interpretation as a two-dimensional object in the dual field theory is still unclear. In addition, one unusual
aspect of our study is the realization that the  $AdS_5\times S^1$ M5-brane probe does not support a self-dual $h$-field. It is ruled out by the equations of motion. 

We have also identified other BPS and non-BPS probes that should correspond to some non-local objects in the dual theory. We hope to study these objects more in future work and provide some better illumination of their properties. As a future direction, one can immediately imagine generalizing the probes we have identified in MN to the more general class of LLM geometries. Another open avenue concerns backreacting the probe branes in the literature along the lines of work pioneered by Lunin for $AdS_4 \times S^7$ and $AdS_7 \times S^4$.

\section*{Acknowledgements}
We are grateful to Nadav Drukker, Bo Feng, Nakwoo Kim, Sungjay Lee, Tianjun Li,
Takuya Okuda, Soo-Jong Rey, Takao Suyama, Satoshi Yamaguchi and Hyun
Seok Yang for useful discussions and comments. JW and HY wish to thank the
Institute of Theoretical Physics (ITP), the Chinese Academy of
Sciences for hospitality during the process of compiling this paper. E\'OC is also grateful to 
the Center of Mathematical Sciences, Zhejiang University for a warm stay and, finally, JW
thanks the School of Physics, Peking University for gracious hospitality. The
work of BC was partially supported by NSFC Grant No.10775002,10975005 and NKBRPC (No. 2006CB805905).
\appendix

\section{Maldacena-N\'u\~nez background}
The ansatz for the d=7 metric appearing in \cite{MN} is \be
ds^{2}_{7} = e^{2 f(r)} (-dt^2 + dx_i^2 + dr^2) + \frac{e^{2
g(r)}}{y^2} (dx^2 + dy^2). \ee From the supersymmetry variations in
\cite{MN}, the general solution where $x,y$ define a Hyperbolic
space may be written \bea e^{5 \lambda} &=& \frac{e^{2 \rho} +
\tfrac{1}{2} + C_1 e^{-2 \rho}}{e^{2 \rho} + \tfrac{1}{4}}, \nn e^{2
g} &=& e^{\lambda} (e^{2 \rho} + \tfrac{1}{4}), \nn e^{2 f} &=& C_2
e^{2 \rho} e^{\lambda}, \nn e^{2 f} \left( \frac{d r}{d \rho}
\right)^2 &=& e^{-4 \lambda}. \eea Here $\rho \rightarrow \infty$
corresponds to the boundary of $AdS_7$, $C_2$ is a trivial
integration constant that may be absorbed by a volume rescaling, and
when $C_1 = 0$, the solution interpolates between $AdS_7$ and $AdS_5
\times \Sigma$. The solution in the IR has the fixed point values
\be e^{5 \lambda } =2, \quad e^{2 g - \lambda} = \frac{1}{4}, \quad
e^{f + 2 \lambda} = \frac{1}{r}, \ee making the d=7 metric \be
ds^{2}_{7} = e^{\lambda} \left[ \frac{1}{2} ds^{2}_{AdS_{5}} +
\frac{1}{4} \frac{(dx^2 + dy^2)}{y^2} \right]. \ee Uplifting this
solution to d=11 makes use of section 4 from \cite{Cvetic:1999xp},
and in particular, the following formula \bea ds^{2}_{11} &=&
\tilde{\Delta}^{1/3} ds^{2}_{7} + g^{-2} \tilde{\Delta}^{-2/3}
\left( X_0^{-1} d \mu_0^2 + \sum_{i=1}^{2} X_i^{-1} ( d \mu_i^2 +
\mu_i^2 ( d \phi_i + g A^{i})^2 )\right), \nn {*}_{11} H_4 &=& 2 g
\sum_{\alpha =0}^{2} \left( X_{\alpha}^2 \mu_{\alpha}^2 -
\tilde{\Delta} X_{\alpha} \right) vol_7 + g \tilde{\Delta} X_0 vol_7
+ \frac{1}{2g} \sum_{\alpha = 0}^{2} X_{\alpha}^{-1} {*}_7 d
X_{\alpha} \wedge d (\mu_{\alpha}^2) \nn &+& \frac{1}{2 g^2}
\sum_{i=1}^{2} X_{i}^{-2} d (\mu_i^2) \wedge (d \phi_i + g A^{i} )
\wedge {*}_{7} F^{i}, \eea where ${*}_{7}$ and $vol_{7}$ are the
Hodge dual and the volume form with respect to the metric $ds^2_{7}$
and ${*}_{11}$ is the Hodge dual of the uplifted metric. In
addition, we have the following relationships: \be X_0 \equiv (X_1
X_2)^{-2}, \quad \tilde{\Delta} = \sum_{\alpha=0}^{2} X_{\alpha}
\mu_{\alpha}^2, \quad \sum_{\alpha = 0}^{2} \mu_{\alpha}^2 =1. \ee
Making contact between the two actions, i.e. (89) of \cite{MN} and
(4.6) of \cite{Cvetic:1999xp}, means adopting the following
identifications \bea g &=& 2, \quad X_0 = X_2 = e^{2 \lambda}, \quad
X_1 = e^{-3 \lambda}, \quad 2 A_{MN}^{i} = A^{i} \nn \mu_0 &=& \cos
\theta \cos \psi, \quad \mu_1 = \sin \theta, \quad \mu_2 = \cos
\theta \sin \psi. \eea With these identifications \be \tilde{\Delta}
= e^{-3 \lambda} (1+ \cos^2 \theta) = e^{-3 \lambda} W. \ee We also
obtain the metric \bea ds^{2}_{11} &=& \frac{1}{2} W^{1/3}
ds^{2}_{AdS_5} + \frac{W^{-2/3}}{4} \biggl[ W \frac{(dx^2 +
dy^2)}{y^2} \nn &+& W d \theta^2 + \cos^2 \theta (d \psi^2 + \sin^2
\psi d \phi_2^2) + 2 \sin^2 \theta \left( d \phi_1 +
\frac{dx}{y}\right)^2\biggr], \eea where we have used $A_{MN}^{1} =
y/4 dx$. One may also determine the fluxes from the formula above.
The term \be 2 g \sum_{\alpha =0}^{2} \left( X_{\alpha}^2
\mu_{\alpha}^2 - \tilde{\Delta} X_{\alpha} \right) vol_7 + g
\tilde{\Delta} X_0 vol_7 \ee gives the following contribution to
$H_4$: \be -\frac{1}{4 W^2} [ 3 + \cos^2 \theta ] \sin \theta \cos^2
\theta d \theta (d \phi_1 + \frac{dx}{y}) \sin \psi d \psi d \phi_2.
\ee The next term is zero and the last term \be \frac{1}{2 g^2}
\sum_{i=1}^{2} X_{i}^{-2} d (\mu_i^2) \wedge (d \phi_i + g A^{i} )
\wedge {*}_{7} F^{i}, \ee becomes \be \frac{1}{4} \frac{\cos^{3}
\theta}{W} \frac{dx dy}{y^2} \sin \psi d \psi d \phi_2. \ee
Up to relabeling of coordinates, this is the form of the solution appearing in the introduction.

\section{MN Killing spinors}

Parallel to the treatment in \cite{LLM}, we introduce a
decomposition for the D=11 gamma matrices satisfying \be
\{\G^{M},\G^{N} \} = 2 \eta^{MN}, \;\; M,N=0,...,10. \ee These may
be re-expressed in terms of lower-dimensional gamma matrices as \bea
\label{gamma_dec} \G^{a} &=& \rho^{a} \otimes \sigma_3 \otimes
\gamma_{(4)}, \nn \G^{i} &=& 1_4 \otimes \sigma_{i} \otimes
\gamma_{(4)}, \nn \G^{\m} &=& 1_4 \otimes 1_2 \otimes \gamma_\m,
\eea where $a=0,...,4$ denote $AdS_5$ directions, $i=1,2$ denote
directions along the $S^2$ and $\m=7,...,10$ label gamma matrices
along the remaining $(x,y,\chi,\theta)$ directions, respectively.
$\g_{(4)}$ is simply the product $\g_{78910}$ and we adopt the sign
choice $\G_{012345678910} = -1$. Consequently, this implies that
$\rho_{01234} = i$.

Throughout this paper, we will make use of the explicit construction of Killing spinors on AdS spacetimes appearing in \cite{Pope}. Writing the $AdS_5$ metric as
\be
ds^{2}_{AdS_5} = \tfrac{1}{r^{2}} \left( dx_{\mu} dx^{\mu} +{dr^2} \right),
\ee
the solutions to the Killing spinor equation
\be \label{ads5_kse} D_{a} \psi = \frac{1}{2} \rho_a \psi,\ee may be expressed as
\bea
\label{poincare} \psi_{+} &=& r^{-1/2} \eta_{+}, \\
\label{sconf} \psi_{-} &=& r^{1/2} \eta_{-} + r^{-1/2} x^{\alpha} \rho_{\alpha} \eta_{-},
\eea
where $a=0,\cdots, 4$ labels the $AdS_5$ coordinates including $r$, and $\a = 0,\cdots, 3$ omits $r$. The constant spinors $\eta_{+}, \eta_{-}$ correspond to Poincar\'{e} and superconformal Killing spinors and are subject to the additional projection condition \cite{Pope}
\be \rho_{r} \eta_{\pm} = \pm \eta_{\pm}. \ee

Note that replacing $\psi$ with its conjugate $\psi^{c}$ results in a sign change in the Killing spinor equation (\ref{ads5_kse}). This knock-on effect of this change is that $\eta_{+}$ and $\eta_{-}$ get interchanged in the solution.

From here on, we write down a general expression for a Killing spinor preserved by the MN background as
\bea
\e &=& \psi \otimes \xi + \psi^{c} \otimes \xi^{c},
\eea
where we just focus on the Poincar\'{e} Killing spinors
\be  \psi = r^{-1/2} \eta_{+}, \quad \psi^{c} =
r^{-1/2} \eta_{-}. \ee

The rest of this section concerns the identification of $\xi$ and $\xi^c$ from the Killing spinor equation for the MN background.

We begin by examining the eleven-dimensional Killing
spinor equation \be \nabla_{m} \eta + \frac{1}{288}
\left[\G_{m}^{~npqr} - 8 \delta_{m}^{~n} \G^{pqr} \right]
H_{4 ~npqr} \eta = 0. \ee As in LLM \cite{LLM}, the analytically continued solution  may be reduced on $AdS_5$, then $S^2$,
before the differential constraints on the remaining
four-dimensional space may be extracted. Within the framework of LLM we can incorporate
the two sign choices in the $AdS_5$ Killing spinor equation as: \be D_a\Psi=(ib)\frac{i}{2}\rho_a\Psi,
\ee where $b=-1$ for $\Psi=\psi$ and $b=1$ for $\Psi=\psi^c$.

Introducing the gamma matrix decomposition introduced earlier (\ref{gamma_dec}) and following the steps as outlined in appendix F of LLM, one arrives at
the following equations \be
\left[\g^\m\partial_\m\lambda+\frac{a}{12}e^{-3\lambda-2A}\g_{(4)}\g^{\m\n}
F^{(2)}_{\m \n}+iabm\right]\epsilon=0, \label{F20} \ee
\be\left[ie^{-A}\g_{(4)}+\g^\m\p_\m
A-\frac{a}{4}e^{-3\lambda-2A}\g_{(4)}\g^{\m\n} F^{(2)}_{\m \n}-iabm
\right]\epsilon=0,\label{F21}\ee \be
\left[\nabla_\m-i\frac{abm}2\g_\m-\frac{a}{4}e^{-3\lambda-2A}F^{(2)}_{\m\n}\g^\n\g_{(4)}
\right]\epsilon=0,\label{F22} \ee with $a = \pm 1$ and $F^{(2)}$
defined in terms of the four-form flux by \be H_{4} = F^{(2)}
\wedge d^{2} \Omega. \ee

As in LLM combining (\ref{F20}) and (\ref{F21}) to remove the
$F^{(2)}$ terms, we find the condition \be
\p_\m(A+3\lambda)\g^\m\epsilon+ie^{-A}\g_{(4)}\epsilon+2iabm\epsilon=0.\label{F35}
\ee Taking $m=\frac12$, the solution to this equation is \be
\label{rotate} \epsilon=e^{-i {a b
\gamma_{10}\phi_0}/{2}}\epsilon_0,\ee where \be \label{phi0}
\sin\phi_0=\frac{\sqrt{2}\cos\theta}{\sqrt{W}},\quad
\cos\phi_0=-\frac{\sin\theta}{\sqrt{W}}, \ee and $\epsilon_0$
satisfies the projection condition \be \label{proj1}
(i\g_{10}\g_{(4)}+1)\epsilon_0=0. \ee

Returning then to (\ref{F20}), we may determine the second
projection condition from the requirement that it is satisfied by
the MN solution. This amounts to the following being satsified
\bea\left[-\frac{\sin\th\cos\th}{3\sqrt{2}W}\g_{10}-\frac{a\cos\th}{3\sqrt{2W}}\g_{9
10}- \frac{a(3+\cos^2\th)}{6W}\g_{78}+iabm\right]\epsilon=0.
\label{201}\eea Using (\ref{rotate}), $m=\frac{1}{2}$, $\gamma_9
\e_0 = \alpha \e_0$\footnote{Implying $i \g_{78} \e_0 = \alpha \e_0$
from (\ref{proj1}).} where $\alpha = \pm 1$, we may expand this
expression to get two parts, one proportional to the identity and
the other proportional to $\g_{10}$: \bea &&
\g_{10}\left[-\frac{sc}{3\sqrt{2}W}(1-\frac{s}{\sqrt{W}})+\frac{c}{\sqrt{2W}}+
\alpha a \frac{c}{3\sqrt{2W}}(1-\frac{s}{\sqrt{W}}) + \alpha b
\frac{(3+c^2)c}{3\sqrt{2}W\sqrt{W}} \right] \e_0 = 0 \nn &&1_{4}
\left[ab\frac{sc^2}{3W\sqrt{W}}+\frac{1}{2}ab(1-\frac{s}{\sqrt{W}})
+ \alpha b \frac{c^2}{3W} + \alpha a
\frac{(3+c^2)}{6W}(1-\frac{s}{\sqrt{W}}) \right] \epsilon_0=0. \eea
Here we have employed the shorthand $c \equiv \cos \th$, $s \equiv
\sin \th$ to compress these expressions. These are satisfied provided \be a = b = - \alpha, \;\;
\mbox{with} \;\; a^2 = 1. \ee

The final part of the Killing spinor may be determined by solving (\ref{F22}) directly to determine the functional dependence of $\e_0$. From the projectors $i \g_{78} \e_0 = \alpha \e_0$ and $\g_9 \e_0 = \alpha \e_0$, we can track the dependence on the sign $\alpha$.   After examining (\ref{F22}) one determines that
\be
\partial_{\theta} \e_0 = \partial_{x} \e_0 = \partial_{y} \e_0 = 0.
\ee
The dependence on $\chi$ may easily be determined from the Killing spinor equation in the $x$ or $\chi$ directions. For ease of illustration we focus on the $x$ direction. After a small calculation this becomes
\bea
\biggl[ \frac{y}{\sqrt{2}} \left( \partial_{x} - \frac{\partial_{\chi}}{y} \right) &+& \frac{1}{2} \left( - \frac{1}{\sqrt{2}} \g_{78} - \frac{\sin \th}{2 \sqrt{W}} \g_{98} \right)  \nn &-& \frac{i a b}{4} \g_7 + \frac{a\sqrt{2} \cos \th}{4 \sqrt{W}} \g_{789} \biggr] e^{-\tfrac{i}{2} {a b
\gamma_{10}\phi_0}}\epsilon_0(\chi) = 0.
\eea
In this expression, after using the projectors, there are terms proportional to $1_4$, $\g_8$ and $\g_{810}$. The latter two cancel independently of their own accord, but the term proportional to the identity becomes
\be
\left[ \partial_{\chi} + \frac{1}{2} \g_{78} \right] \e_0(\chi) = 0.
\ee
Using $i \g_{78} \e_0 = \alpha \e_0$, $\e_0(\chi)$ may be written simply as
\be
\e_0 (\chi) = e^{\tfrac{i}{2} \alpha \chi } \tilde{\e}_0,
\ee
where $\tilde{\e}_0$ is a constant spinor.
We can therefore confirm the form of the
Killing spinors for MN appearing in the text (\ref{kills}). The extra $\g_7$ has been introduced to ensure that the projection conditions are correct.

\section{Some connections}
In this appendix, we list the connections appear in
section~\ref{nonbps}. The nontrivial independent components of the
Levi-Civita connection for the induced metric eq.~(\ref{induced})
are: \bea
\Gamma^\zeta_{\tau\tau}&=&-\G^\zeta_{\varphi\varphi}=\cosh\zeta\sinh\zeta,
\\ \G^{\tau}_{\tau\zeta}&=&\frac{\sinh\zeta}{\cosh\zeta}, \\
\G^{\varphi}_{\varphi\zeta}&=&\frac{\cosh\zeta}{\sinh\zeta},\\
\G^x_{yx}&=&\G^y_{yy}=-\frac1y,\\
\G^y_{xx}&=&\frac1y(1+\frac{2\sin^2\th_0}{W_0}). \eea

Some of the nonzero independent components of the spin connection
with respective to the vielbeins in that section are:
\bea\omega^{\underline{1}}_{\underline{0}\underline{0}}&=&\frac{\sqrt{2}\sinh\zeta}{\tilde\kappa^{1/3}W^{1/6}\cosh\rho\cosh\zeta},\\
\omega^{\underline{1}}_{\underline{2}\underline{2}}&=&-\frac{\sqrt{2}\cosh\zeta}{\tilde\kappa^{1/3}W^{1/6}\cosh\rho\sinh\zeta},\\
\omega^{\underline{3}}_{\underline{0}\underline{0}}&=&-\omega^{\underline{3}}_{\underline{1}\underline{1}}=
-\omega^{\underline{3}}_{\underline{2}\underline{2}}=\frac{\sqrt{2}\sinh\rho}{\tilde\kappa^{1/3}W^{1/6}\cosh\rho},\\
\omega^{\underline{3}}_{\underline{4}\underline{4}}&=&-\frac{\sqrt{2}\cosh\rho}{\tilde\kappa^{1/3}W^{1/6}\sinh\rho},\\
\omega^{\underline{8}}_{\underline{7}\underline{7}}&=&\frac2{\tilde\kappa^{1/3}W^{1/6}},\\
\omega^{\underline{10}}_{\underline{0}\underline{0}}&=&-\omega^{\underline{10}}_{\underline{i}\underline{i}}=
-\frac{2\sin\th\cos\th}{3\tilde\kappa^{1/3}W^{1/6}(1+\cos^2\th)},
i=1,\cdots, 4, 7, 8,\\
\omega^{\underline{10}}_{\underline{5}\underline{5}}&=&\omega^{\underline{10}}_{\underline{6}\underline{6}}=
\frac2{\tilde\kappa^{1/3}W^{1/6}}\left(\frac{\sin\th}{\cos\th}-\frac{2\sin\th\cos\th}
{3(1+\cos^2\th)}\right),\\
\omega^{\underline{10}}_{\underline{9}\underline{9}}&=&\frac2{\tilde\kappa^{1/3}W^{1/6}}
\left(-\frac{\cos\th}{\sin\th}-\frac{2\sin\th\cos\th}
{3(1+\cos^2\th)}\right). \eea The remaining non-vanishing components
are: \be \omega^{\underline{5}}_{\underline{6}\underline{6}},
\omega^{\underline{7}}_{\underline{9}\underline{8}},
\omega^{\underline{7}}_{\underline{8}\underline{9}},
\omega^{\underline{8}}_{\underline{7}\underline{9}}, \ee but they
are not needed in the computations in this paper.

\end{document}